\documentclass[fleqn,usenatbib]{mnras}

\usepackage[T1]{fontenc}
\usepackage{newtxtext, newtxmath}

\usepackage{graphicx}
\usepackage{rotating}
\usepackage{tabularx}
\usepackage{ragged2e}
\usepackage{multirow}
\usepackage{enumitem}

\newenvironment{tightitemize}{
  \vspace*{-0.6\baselineskip}
  \begin{itemize}[nosep, leftmargin=*, topsep=0pt, partopsep=0pt, itemsep=0pt, after=\vspace{-\baselineskip}]
}{
  \end{itemize}
  \vspace*{-0.8\baselineskip}
}

\usepackage{fancyhdr}
\title[Human-AI Agent Interaction]{Human-AI Agent Interaction in a Business Context}

\author[K. Paimann et al.]{
Kathrin Paimann,$^{1}$\thanks{E-mail: kathrin.paimann@sap.com}
Elizangela Valarini,$^{1,2}$ and
Sebastian Juhl$^{1,3}$
\\
$^{1}$SAP SE, Dietmar-Hopp-Allee 16, 69190 Walldorf, Germany\\
$^{2}$Hochschule Fresenius Heidelberg, Sickingenstra\ss{}e 63-65, 69126 Heidelberg, Germany\\
$^{3}$University of Missouri, Columbia, USA
}

\date{Version: \today\\
Note: This article is part of the HCI International 2026 Conference Proceedings and will be published in a volume of the Lecture Notes in Computer Science (LNCS) series by Springer Nature.}

\pubyear{2026}

\begin{document}
\label{firstpage}
\pagerange{\pageref{firstpage}--\pageref{lastpage}}
\maketitle

\begin{abstract}
As AI agents are increasingly integrated into core business processes, understanding and designing effective interaction patterns between humans and AI agents becomes crucial for value creation. This study identifies and evaluates principles and criteria for a positive User Experience (UX) with AI agents, along with methods for its measurement. We identify user expectations and needs to facilitate adoption, build trust, and support user-centered decision-making by development teams. Using a mixed-methods approach that combines qualitative and quantitative techniques, we explore interaction patterns between humans and AI agents. The findings from this exploratory research serve as the basis to develop a survey experiment which evaluates the effectiveness of specific design elements on a larger scale. This foundational research contributes to the development of more intuitive and effective human-AI agent interactions in business settings.
\end{abstract}


% remove header from first page
\renewcommand{\headrulewidth}{0pt}
\pagestyle{fancy}
\fancyhead{}
\fancyhead[LE]{\Large \thepage\hspace{.5cm} \textit{K. Paimann et al.}}
\fancyhead[RO]{\Large \textit{Human-AI Agent Interaction} \hspace{.5cm} \thepage}
\fancyfoot{}

\thispagestyle{empty}

%%%%%%%%%%%%%%%%%%%%%%%%%%%%%%%%%%%%%%%%%%%%%%%%%%
%%%%%%%%%%%%%%%%% BODY OF PAPER %%%%%%%%%%%%%%%%%%
\section{Introduction}
Advances in artificial intelligence (AI), particularly in machine learning (ML), natural language processing, and large language models, have driven a transition from static, tool-based automation towards increasingly autonomous, interactive, and context-aware AI systems. The most prominent expression of this transformation in enterprise contexts is the emergence of agentic AI: Autonomous, goal-directed software systems capable of planning multi-step actions, perceiving their operational context, making decisions, and interacting with human counterparts in natural language \citep{bandi2025}. Unlike earlier AI tools, which operated as passive instruments under clear human direction, agentic systems exhibit properties traditionally associated with social actors and occupy organizational roles previously held exclusively by humans \citep{calvanese2026,borghoff2025}. This paradigm shift fundamentally alters the conditions under which business users engage with AI systems and, consequently, the experience they expect from them. Conventional User Experience (UX) practice was developed for non-agentic systems producing predictable, human-controlled outcomes, and proves insufficient to account for the complexity of UX introduced by autonomous, adaptive AI agents.

Although a rapidly expanding body of literature has begun to reconceptualize UX design for the AI era, emphasizing controllability, explainability, ethics, privacy, transparency, and collaborative human-AI interaction as primary design imperatives \citep{li2026,xu2025,xu2024,diederich2022}, a systematic, empirically validated framework of UX principles and measurable criteria applicable specifically to human-AI agent interactions in business contexts has yet to be established.

The present study addresses this gap by investigating how human-AI agent interactions should be designed to satisfy user requirements and needs in business contexts. Implementing a mixed-methods research design, comprising a secondary meta-analysis, a participatory design workshop, a paper-and-pencil survey, semi-structured in-depth interviews, and a conjoint experiment, the study (1) develops a framework and identifies the most critical UX principles for positive human-AI agent interaction in business contexts; (2) derives measurable criteria for each principle; and (3) provides evidence-based design guidelines to support design decisions for the development of AI business agents.

\section{Literature Review and Related Work}
Early Human-Computer Interaction (HCI) and UX research focused primarily on non-intelligent systems, emphasizing usability, interface consistency, and task efficiency. UX practice has since evolved across three eras: UX 1.0 (PC/Internet), UX 2.0 (mobile), and the emerging UX 3.0 paradigm for the AI. This is characterized by ecosystem-wide, AI-enabled experiences spanning products, organizations, and socio-technical environments \citep{xu2024}. Within this evolution, human-AI interaction constitutes a distinct design challenge: AI systems exhibit autonomous, adaptive, and sometimes opaque behavior, introducing risks related to trust, controllability, fairness, and safety that traditional UX methods only partially address \citep{xu2025,shneiderman2020}.

A foundational framework for Human-Centered AI (HCAI) argues that AI systems should augment human capabilities through reliable, safe, and trustworthy design, with explicit principles for human control, transparency, and accountability \citep{shneiderman2020}. Building on this, the UX 3.0 paradigm identifies four categories of emerging experience: Ecosystem-based, innovation-enabled, AI-enabled, and human-AI interaction-based, in which foregrounding explainability, controllability, and ethical alignment is included \citep{xu2024}. The HCAI Methodological Framework (HCAI-MF) operationalizes these principles through a requirement hierarchy linking abstract values to concrete product requirements, an approach-and-method taxonomy, an integrated HCAI process, an interdisciplinary collaboration model, and multi-level design paradigms \citep{xu2025}. While these frameworks provide rich conceptual structure, they offer relatively few empirically validated design criteria tailored to business-critical agent use cases.

\citet{amershi2019} consolidate more than 150 AI-related design recommendations into 18 guidelines organized around four temporal phases: Initial interaction, interaction in context, error handling, and long-term interaction. Validated across 20 commercial AI-infused products, these guidelines codify cross-cutting UX concerns specific to AI, including managing uncertainty and enabling control over adaptive behavior. However, their evaluation focuses on consumer-facing products rather than enterprise workflows, limiting direct translation into quantitative criteria for organizational AI agent use. Explainable AI research similarly reinforces intelligibility as a core design requirement, though empirical evidence remains mixed: poorly aligned explanations can increase overreliance or fail to improve decision quality, indicating that design principles must address socio-organizational conditions of use, not only interface features \citep{xu2025,shneiderman2020}.

\citet{punzi2026} survey hybrid decision-making paradigms and identify three families: Human oversight, learning to abstain, and learning together. These paradigms inform human-in-the-loop (HITL) architectures, such as escalation from AI agent to human expert, and point towards bidirectional, co-creative interaction patterns. Nonetheless, most studies are conducted under experimental conditions with limited, and systematic evidence and validity on how these paradigms affect productivity, accountability, and UX in enterprise workflows remains limited. Complementing these frameworks, \citet{newman2023} and also the \citet{NIST2023} (NIST) address the measurement and governance of trustworthy AI, organizing attributes such as reliability, robustness, fairness, privacy, safety, and ac-countability into structured frameworks for design and risk management. Yet most work at this level targets organizational governance models rather than concrete interaction patterns, and empirical evidence on how trust-related interventions, such as confidence indicators or audit trails, shape user behavior in enterprise contexts remains limited \citep{xu2025,newman2023}. Both HCAI-MF and UX 3.0 advocate multi-layered measurement approaches connecting UX indicators, operational metrics, and organizational outcomes, but standardized evaluation instruments remain emergent \citep{xu2025,xu2024}.

The literature reveals five recurrent gaps: Existing frameworks are predominantly conceptual with few empirically validated criteria \citep{xu2025,xu2024}; guidelines focus on consumer-facing AI rather than enterprise workflows \citep{punzi2026,amershi2019}; quantitative measurement of human-AI interaction remains under-specified \citep{xu2025,punzi2026}; socio-technical and organizational factors are rarely studied empirically \citep{xu2025}; and UX-oriented, ML-oriented, and governance-oriented perspectives remain fragmented \citep{newman2023,shneiderman2020}.

Against this background, the present study contributes to the literature by identifying and structuring the UX principles most critical to human-AI agent interaction in business environment, developing an empirical validated framework, measurable criteria for UX-AI agent’s principles, and providing a foundation for evidence-based design of AI agents in business contexts.

\section{Research Design, Methods, and Sample}
Applying a mixed-methods approach, we collect and analyze empirical data to identify and validate the core UX principles for positive human-AI agent interaction, as well as to translate them into design principles and guidelines.

Combining qualitative and quantitative techniques allows us to derive a holistic understanding of the complex, multidimensional concept of human-AI agent interaction and to provide reliable results that can be utilized to inform practical design decisions in a business context. Specifically, this approach allows us to (1) identify relevant interaction requirements, (2) refine them into design principles, and (3) validate them empirically. Table~\ref{tab:overview} provides an overview over the different methods used in this study as well as the sample.

\begin{table}
  \centering
\rotatebox{90}{
\setlength{\tabcolsep}{4pt}
\scriptsize
\begin{tabular}{p{1.1cm}p{1.9cm}p{1.9cm}p{1.9cm}p{1.9cm}p{1.9cm}p{1.9cm}}
\hline
{\bfseries Method} & {\bfseries Participatory Design Workshop} & {\bfseries Expert Review} & {\bfseries Meta-Analysis} & {\bfseries Paper-and-Pencil Survey} & {\bfseries In-depth Interviews} & {\bfseries Conjoint Experiment}\\
\hline
Sample & 21 & 5 & 28 & 22 & 12 & 107\\
Role & Developers, Software Architects, Product Owners, Data Scientists & AI Experts in Software Design, Research, Development & Research articles & Developers, Software Architects, Product Owners, Data Scientists & Developers, Software Architects, Product Managers, Data Scientists & Developers, Software Architects, Product Mangers, Software Consultants, Business Analysts, UX Designers \\
Industry & Manufacturing, IT Services \& Consulting, Pharmaceutical, Chemical, Wholesale, Insurance & IT Software & IEEE, Academia, ResearchGate, ArXiv, Acm, NIST, MDPI & Manufacturing, IT Services \& Consulting, Pharmaceutical, Chemical, Wholesale, Insurance & IT Software, Healthcare, Manufacturing, Consulting, Retail, Pharmaceutical, E-Commerce & IT Services \& Consulting, Telecommunication, Healthcare, Fintech, Banking, Retail, Utilities, Automotive, Manufacturing, Higher Education
\\
Session Type & In-person moderated & Online unmoderated & Online & In-person moderated & Online Moderated & Online Unmoderated\\
Objective & Explore criteria for positive user experiences with agents. Identify expectations, requirements, and perceived risks. & Validate and consolidate gathered UX criteria for human-agent interaction. & Create a framework mapping the core UX principles and criteria for human-AI agent interaction. & Validate the UX principles and criteria. Prioritize them according to their importance. & Understand prioritization reasons. Explore user specific needs in business environments. Understand the meaning and importance of the prioritized principles and criteria. & Estimate the causal impact of the UX criteria on user preferences.\\
\hline
\end{tabular}
}
\caption{Overview of the research design, applied methods, and sample.}\label{tab:overview}
\end{table}

\subsection{Participatory Design Workshop}
The participatory design workshop is employed to explore early principles and criteria—from the perspectives of domain experts and end users—that contribute to a positive user experience when working with AI agents in a business context. The workshop titled ``The Future of Work with AI Agents'' additionally serves to identify needs, requirements, expectations, and perceived risks across participant groups. It was conducted in-person with participants (self-)selected\footnote{The participatory design workshop was conducted as part of a UX-organized event, and participants registered voluntarily.} on the basis of prior experience with AI systems and familiarity with agent-based technologies. Participatory design workshops are most commonly applied in software development contexts \citep{muller1993,schuler1993,greenbaum1991}, where stakeholders, including end-users, domain experts, and design practitioners are convened in structured, facilitated sessions to collaboratively address a defined problem, such as articulating needs, and generating design propositions \citep{wacnik2025}. Over the course of a 60-minute session, participants drawn from diverse industry sectors and occupying central roles in the software development lifecycle were organized into smaller groups according to role similarity, with each group tasked with exploring agent use cases relevant to their domain and identifying criteria that would ensure a positive interaction experience with agents. Participants are positioned as active co-producers of the research outcome, as their tacit knowledge constitutes a primary epistemic source in the absence of a pre-existing framework for this domain \citep{arcia2024}.

\subsection{Expert Review and Meta-Analysis}
To cluster the findings of the participatory design workshop regarding the explored UX principles and criteria for positive user experience in human-AI agent interaction, five domain experts in AI review the findings and identify recurring topics. Following this step, a qualitative meta-analysis serves as the methodological foundation to aggregate and interpret existing findings from heterogeneous studies through analytical steps \citep{paul2022,ralph2022,lachal2017,sandelowski2007}. This approach is particularly suited in this context, since the existing literature is rich in UX conceptual frameworks and UX criteria but lacks an integrative conceptual structure. Furthermore, the absence of an established domain-specific framework renders primary data collection alone an insufficient basis for theory-building at this stage of research. A corpus of peer-reviewed publications and a relevant standards framework (see Table~\ref{tab:corpus}) was therefore analyzed to develop a framework consisting of eight core UX principles and underlying criteria for positive user experience in the context of human-AI agent interaction (see Table~\ref{tab:principles}). Keywords used for the selection of the corpus included HCAI, HCI, HAI, agents, frameworks, principles, and business or organizational context. Based on our research question, data are synthesized to derive the higher-order categories and criteria \citep{shen2024,sandelowski2007}, consistent with established practice in qualitative meta-analysis for framework development, in which the synthesized body of literature functions as the primary data source and the resulting framework constitutes an original theoretical contribution \citep{ralph2022}.

\begin{table*}
  \centering
\setlength{\tabcolsep}{4pt}
\setlength{\extrarowheight}{4pt}
\begin{tabular}{p{3.2cm}p{5cm}p{1cm}p{6cm}}
\hline
{\bfseries Authors} & {\bfseries Title } & {\bfseries Year} & {\bfseries Main Contribution}\\
\hline
Bradshaw, J.M., Feltovich, P., Johnson, M. & Human-Agent Interaction & 2011 & Defines joint human-agent activity and key design requirements. Describes maxims for a good agent.\\
Diederich, S., Brendel, A.F., Morana, S., Kolbe, L. & On the Design of and Interaction with Conversational Agents: An Organizing and Assessing Review of Human-Computer Interaction Research & 2022 & HCI review of conversational agents, agent interaction design, and UX outcomes in applied organizational contexts. \\
Shneiderman, B. & Human-Centered Artificial Intelligence: Reliable, Safe \& Trustworthy & 2020 & HCAI reference for reliability, safety, trustworthiness, and human oversight in AI-supported work.\\
Xu, W. & A User Experience 3.0 Paradigm Framework/ User Experience 3.0 Paradigm Framework & 2024 & UX 3.0 framework translating AI-specific concerns such as trust, transparency, and control into actionable UX dimensions.\\
Xu, W., Gao, Z., Dainoff, M. & An HCAI Methodological Framework (HCAI-MF): Putting It Into Action to Enable Human-Centered AI & 2025 & Operational HCAI framework with guiding principles including human control, UX, transparency, safety, and accountability.\\
NIST & AI Risk Management Framework 1.0 & 2023 & Governance framework linking human oversight, transparency, robustness, and risk management in organizational AI use.\\
Nastoska , A., Jancheska, B., Rizinksi, M., Trajanov, D. & Evaluating Trustworthiness in AI Risks, Metrics, and Applications Across Industries & 2025 & Business-relevant review of trustworthy AI across industries including finance, healthcare, and public administration; selected for linking trust, governance, and applied organizational adoption.\\
\hline
\end{tabular}
\caption{Corpus of selected articles and journals.}\label{tab:corpus}
\end{table*}

\begin{table}
  \centering
  \setlength{\tabcolsep}{4pt}
  \tiny
  \rotatebox{90}{
    \begin{tabular}{p{1.2cm}p{2.3cm}p{2.3cm}p{2.3cm}p{2.3cm}p{2.3cm}p{2.3cm}p{2.3cm}p{2.3cm}}
      \hline
      {\bfseries UX Principle} & 
      {\bfseries A human is always in control of an agent.} & 
      {\bfseries An agent operates transparently and explains its output.} & 
      {\bfseries An agent is reliable, safe and robust.} & 
      {\bfseries An agent is always context-aware.} & 
      {\bfseries An agent acts as a collaborative partner.} & 
      {\bfseries An agent adheres to data privacy \& data governance.} & 
      {\bfseries An agent is integrated into the ecosystem.} & 
      {\bfseries An agent is responsive and intuitive to use.} \\
      \hline
      
      UX Criteria & 
      A human maintains ultimate authority, with the ability to intervene, override, or disengage the agent. & 
      An agent provides clear reasoning and describes ``how'' and ``why'' it came to a result. & 
      An agent’s output is accurate and avoids hallucinations or misleading information. & 
      An agent is aware of the user's role and permissions when providing relevant assistance. & 
      A human primarily interacts with a single ``orchestrating'' agent that coordinates a team of specialized agents and reroutes queries.& 
      An agent enforces strict data privacy protection. & 
      An agent has access to the full ecosystem of enterprise applications, including sap and third-party systems. & 
      An agent can be used seamlessly across devices. \\
      \\
      
      & 
      A human is always responsible for the decisions made in relation to an agent's actions. & 
      An agent's actions and outputs can be easily verified. & 
      An agent always uses up-to-date resources and data. & 
      An agent provides guidance and support for multi-stage processes or tasks, maintaining context from one step to another. & 
      A human can add and remove specialized agents from an interaction, task or process. & 
      An agent enforces role-based access controls to ensure the output is appropriate for the user's permissions. & 
      An agent functions as a collaborative process orchestrator to manage complex workflows from start to finish.
      The primary interaction between the human and agent is a conversational UI that adheres to UX best practices (i.e., accessibility, consistency, intuitiveness etc.). & \\
      \\
      
      & 
      A human always has to confirm before an agent takes a critical (business) decision. & 
      An agent’s output is understandable for the user's specific skill level and context. & 
      An agent provides clear risk indications (e.g., knowledge limits or uncertainty). & 
      An agent understands questions in the context of recently uploaded data, enabling it to provide accurate answers and relevant suggestions. & 
      A human is always aware of all specialized agents involved in an interaction even if s/he does not directly interact with them. & 
      Humans provide consent over how their data is collected and used. & 
      An agentic system can access multiple apps seamlessly and abstract background complexity into a simple user interface. & 
      An agent communicates in natural language. \\
      \\
      
      & 
      A human is always aware of the agent's intended steps or actions before it executes them. & 
      An agent provides sufficient information on how it works, including the rules or logic it follows. & 
      An agent clearly shows which human input is needed (e.g., decision). & 
      An agent adapts over time to the personal working style of a user. & 
      An ``orchestrating'' agent acts like a ``teammate'' that performs tasks while asking for approval or clarification at key An agent operates predictably by adhering to data governance protocols. & 
      Predictable through governance protocols. & 
      & 
      A human can easily determine an agent's current status and progress. \\
      \\
      
      & 
      An agent works autonomously only on low-impact, time-consuming and/or manually intensive tasks. & 
      A human can easily determine an agent's current status and progress. & 
      When an agent fails to understand an input, it applies ``repair strategies'' that are actionable and explanatory.& 
      An agent is aware of the user's team structure and context. & 
      & 
      An agent follows robust security standards to protect the system and the data it accesses. & 
      &
      \\
      \\
      
      & 
      & 
      An agent provides completion reports to confirm accomplishment.& 
      An agent implements audit trails to reveal its process (e.g., activity log). & 
      & 
      & 
      & 
      & 
      \\
      \hline
    \end{tabular}
  }
\caption{Identified UX principles and criteria.}\label{tab:principles}
\end{table}

\subsection{Paper-and-Pencil Survey}
The constructed framework, comprising eight core UX principles and three to six criteria for positive user experience in human-AI agent interaction in a business context, serves as the basis for conceptualizing a paper-and-pencil survey. This method aims to validate the framework and prioritize the most important UX principles and corresponding criteria based on end-user and expert requirements. As a data collection instrument embedded within the broader mixed-methods design of this study, the survey also quantifies findings from the exploratory phase \citep{boynton2004,burns2008}. The self-administered survey was fielded during the workshop ``Human-AI Agent Interaction'', enabling participants to record independent responses to survey items immediately following a collaborative session activity, with 45 minutes allocated for survey completion and 15 minutes for a subsequent group activity. The self-administered paper format is selected for its methodological suitability in supervised, in-person settings: It standardizes data collection by presenting all participants with identical items in a uniform sequence, eliminates facilitator influence on response content, and has been shown empirically to produce fewer missing responses and fewer extreme response patterns than computerized equivalents under equivalent conditions \citep{edwards2024,colasante2019}. Participants supplemented the paper-and-pencil survey with optional written feedback and discussed their rankings in role-specific groups.

\subsection{In-Depth Interviews}
In the present study, 12 semi-structured in-depth interviews were conducted to deepen the understanding of participants’ prioritization of the UX principles identified through the preceding survey. The interview sessions were conducted virtually, with participants from the paper-and-pencil workshop, a recruiting platform, and inhouse. All participants had experience with AI tools, familiarity with AI agents, and a professional business context. The semi-structured format enables participants to articulate the reasoning, values, and contextual factors that underlie their judgments \citep{mcintosh2015}. The qualitative data serves to contextualize and explain findings from the prior survey phase. The interview guide consists of a small set of core questions focused on participants’ top three ranked UX principles and criteria for human-AI agent interaction, organized to progress from broader, context-setting questions to more specific and in-depth probes. The flexible structure of the guide further allowed for follow-up on emergent themes, enabling a more nuanced exploration of the phenomenon under investigation \citep{dejonckheere2019,kvale1996}. The data were analyzed using thematic analysis following a deductive-inductive approach, in which an initial codification of the material was structured according to the pre-defined interview categories and subsequently complemented by newly emerged themes identified during analysis \citep{braun2006}.

\subsection{Conjoint Experiment}
In a next step, we aim to provide specific design guidance and quantify their effects on the users by implementing a choice-based conjoint experiment in an online survey. Introduced in the 1970s, conjoint experiments allow researchers to simultaneously estimate the causal effect of different treatment variables on the expressed preferences of the respondents \citep{green1971}. Since the setup allows researchers to specify scenarios that closely mirror real-world settings, this research design enhances the external validity of the findings. Together with the ability to establish causality due to the randomization, these appealing characteristics have led to an increase in the application of this survey technique across the social sciences \citep{liu2023,bansak2018,hainmueller2014}.

As the conjoint experiment aims to quantify the effects of UX criteria on users when these are translated into interface design, the instrument needs to be developed based on the triangulated findings from the exploratory research phase, comprising the paper-and-pencil survey and the semi-structured in-depth interviews.

\section{Results}
\subsection{Exploratory Research and Framework Validation}
The paper-and-pencil survey validates the framework, which comprises eight UX principles for human-AI agent interaction and their underlying criteria (see Table~\ref{tab:principles}). No additional principles were identified, and none of the proposed principles was rated as irrelevant. By asking participants to rank the principles according to their importance in a business context, the survey identified the following top three priorities: Human Control (Top 1), Reliability, Safety, and Robustness (Top 2), Data Privacy and Governance, and Context-Awareness (both Top 3).

\begin{table*}
  \centering
\setlength{\tabcolsep}{4pt}
%\scriptsize
\begin{tabular}{p{2.5cm}p{2.8cm}p{5.5cm}p{5.5cm}}
\hline
{\bfseries Principle name} & {\bfseries Priority } & {\bfseries Top UX criteria} & {\bfseries UX design transfer ideas}\\
\hline
Human Control &
\begin{tightitemize}
\item Highest priority
\item 65\% Top 3
\item 40\% Ranked Top 1
\end{tightitemize}&
\begin{tightitemize}
\item 95\% Human confirms critical decisions
\item 86\% User can intervene, override, or disengage
\item 40\% Human remains accountable 
\end{tightitemize}&
Human sign off for high-impact actions; pause/stop controls; accountability advice.\\
\hline

Reliability, safety, robustness &
\begin{tightitemize}
\item $2^{nd}$ overall
\item 60\% Top 3
\end{tightitemize}&
\begin{tightitemize}
\item 65\% Accurate output without hallucinations
\item 57\% Up-to-date sources
\item 57\% Risk indicators
\end{tightitemize}&
Confidence indicators for accuracy; source recency labels; risk or uncertainty flags.\\
\hline

Data privacy and governance & 
\begin{tightitemize}
\item 45\% Top 3
\end{tightitemize}&
\begin{tightitemize}
\item 85\% Role-based access control
\item 76\% Strict privacy protection 
\item 76\% Robust security standards
\end{tightitemize}&
Permission-aware outputs; deny-and-explain for restricted requests; sensitivity labels when sensitive data is accessed.\\
\hline

Context-awareness &
\begin{tightitemize}
\item 45\% Top 3
\end{tightitemize}&
\begin{tightitemize}
\item 95\% Aware of user role and per-missions for relevant assistance
\item 72\% Maintains context in multi-step tasks
\item 72\% Uses recently uploaded data appropriately
\end{tightitemize}&
Role-aware assistance; persistent task context; proactive clarification when context is missing.\\
\hline

Transparency and explainability &
\begin{tightitemize}
\item 40\% Top 3
\end{tightitemize}&
\begin{tightitemize}
\item 75\% Verifiable actions and outputs 
\item 60\% Clear reasoning 
\item 55\% Explanation of rules or logic
\end{tightitemize}&
Source traceability; clear human vs. agent output indicators; ``why/ how'' explanations with expandable detail based on task risk and user need.\\
\hline

Ecosystem integration &
\begin{tightitemize}
\item 30\% Top 3
\end{tightitemize}&
\begin{tightitemize}
\item 40\% Seamless access across applications
\item 40\% Abstract complexity
\item 31\% Orchestration across end-to-end workflows 
\end{tightitemize}&
Cross-tool integration showing active tool usage; use progressive disclosure to surface complexity only when needed; visualize agent actions and handoffs across systems.\\
\hline

Collaborative partner &
\begin{tightitemize}
\item 15\% Top 3
\end{tightitemize}&
\begin{tightitemize}
\item 95\% Add/remove specialized agents 
\item 85\% Orchestrating agent acts as teammate with approval moments 
\item 75\% Visibility of active agents 
\end{tightitemize}&
Allow users to add or remove agents from tasks or processes; clearly identify agents; Multi-agent visibility.\\
\hline

Responsiveness and intuitiveness &
\begin{tightitemize}
\item Lower relative priority, but basic adoption 
\end{tightitemize}&
\begin{tightitemize}
\item 35\% High-quality conversational interaction
\item 30\% Clear status and progress visibility
\end{tightitemize}&
Conversational UI aligned with core UX best practices; clear status indicators and progress bars; use visual representations where applicable.\\
\hline
\end{tabular}
\caption{Ranked UX principles and criteria with design transfer recommendations.}\label{tab:ranked}
\end{table*}

The results clearly show that a positive experience when interacting with an agent depends primarily on maintaining meaningful human control, ensuring reliable and governed agent behavior, grounding outputs in user and task context, and making agent actions transparent enough to support verification without overloading the interaction. The analysis of the in-depth interviews indicated that all eight principles were evaluated in relation to organizational risk, accountability, and practical workflow requirements.

Interviewees understand ``Human Control'' as meaningful oversight at critical decision points, emphasizing final approval, confirmation workflows, and stop mechanisms. ``Reliability, Safety, and Robustness'' are treated as a non-negotiable baseline, as misleading or unstable outputs were seen as disruptive and trust-eroding. ``Data Privacy and Governance'' are rated highly important, particularly in regulated or client-facing contexts, whereas participants noted that access restrictions must remain role-sensitive. ``Context-Awareness'' emerges as a distinguishing feature of a useful enterprise assistant: Agent outputs must reflect the user's role, tasks, and current context. ``Transparency'' is valued as a basis for trust and verification, though its importance varies with task complexity and experience level. ``Ecosystem Integration'' is considered essential for avoiding manual work and enabling seamless workflows, while ``Collaboration as a Partner'' is appreciated when it improves efficiency without reducing oversight. ``Responsiveness and Intuitiveness'' are treated as basic usability requirements rather than differentiating features, as poor interaction quality would undermine adoption regardless of output quality.

The results further indicate that transparency needs may decrease as trust increases, that privacy expectations vary by role and use case, and that multi-agent collaboration requires visibility across all involved agents to prevent loss of oversight. Table~\ref{tab:ranked} presents the eight validated UX principles for human-AI agent interaction and their ranking by importance in a business context.

\subsection{Evaluative Research: Conjoint Study}
Based on the triangulation of the paper-and-pencil survey and the qualitative interviews, we develop design prototype screens to assess the effectiveness of specific design elements. To keep the volume manageable, we restrict the focus on the UX principle ``Human Control'' (see Table~\ref{tab:principles}) which was identified to be the most relevant principle in the paper-and-pencil survey and in our in-depth interviews. Specifically, we operationalize the following three UX criteria for Human Control which are not mutually exclusive: 
\begin{itemize}
\item {\bfseries Ability to stop/pause agents:} Participants consistently emphasized the need to interrupt agent execution, reflecting a fundamental requirement for human authority over automated processes.
\item {\bfseries Transparency of agent reasoning:} Participants required visibility into agent reasoning and actions. This criterion is inherently linked to the first; users need to understand what is happening before they can intervene.
\item {\bfseries Human accountability and decision-making:} Participants indicated a clear need to remain responsible for decisions, particularly in business contexts where audit trails are required.
\end{itemize}

Table~\ref{tab:conjoint} summarizes the different levels each of the UX criteria can take. Together, we develop a total of 12 prototype screens with all possible level combinations. The screens are designed such that they closely mirror a real-world business scenario of a logistics employee responsible for shipping operations and monitoring using an AI agent to support with optimal truck loading. This hypothetical scenario and the developed prototype screens closely mirror a real business use case which ensures the external validity of the results.

\begin{table}
  \centering
\setlength{\tabcolsep}{4pt}
\setlength{\extrarowheight}{4pt}
\begin{tabular}{p{1.5cm}p{2cm}p{2cm}p{2cm}}
\hline
UX Criteria & \multicolumn{3}{c}{Levels}\\
\hline
Stop/pause agent actions & \textit{Low:} Absent & & \textit{High:} Present\\
Agent transparency & \textit{Low:} Simple loading indicator without any additional information. & \textit{Medium:} Display tasks and current status but no detailed reasoning or data sources & \textit{High:} Detailed information on status, next steps and  data sources as well as reasoning and a `View Details' option\\
Human accountability and decision-making & \textit{Low:} Standard disclaimer without warning message or accountability indicator & & \textit{High:} Standard disclaimer plus warning message for `Low Confidence' recommendations\\
\hline
\end{tabular}
\caption{Attribute levels of the UX criteria belonging to the UX principle human control.}\label{tab:conjoint}
\end{table}

Overall, the analysis is based on a sample of $107$ respondents which have completed a total of $1,139$ choice tasks. The respondents are recruited SAP internally and externally via a recruiting platform. All respondents have experience with AI usage in a business context and are familiar with use and/or creation of AI agents. For the analysis, we focus on estimating the average marginal component effect (AMCE) of each level of the UX criteria under consideration. This quantity is the causal marginal effect of an UX criterion and can be interpreted as the change in the probability that the prototype screen is selected if the UX criterion changes its level, averaged over the remaining UX criteria \citep{bansak2023,hainmueller2014}. Since the respondents complete binary choice tasks, we estimate a conditional logistic regression model to compute AMCEs. We also use clustered standard errors to account for the fact that respondents perform multiple choice tasks \citep{lehrer2024}.

\begin{figure}
\centering
\includegraphics[width=.49\textwidth]{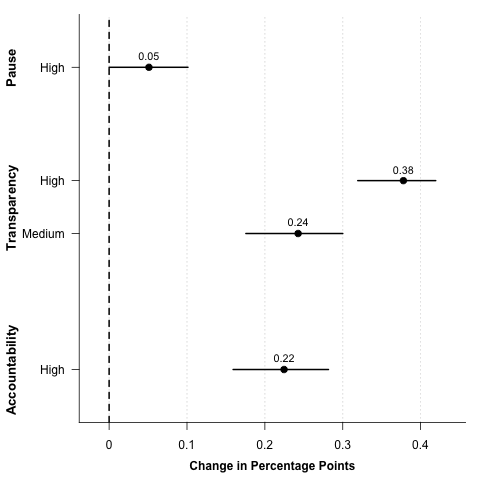}
\caption{Average Marginal Component Effects of the UX criteria related to the human control UX principle.} \label{fig:amces}
\end{figure}

Figure~\ref{fig:amces} visually presents the AMCEs which can be interpreted as the expected change in the probability to prefer a given screen prototype compared to a reference screen which we define as the prototype screen where the three UX criteria under consideration takes on the lowest level. The points represent the point estimate of the AMCEs while the bars indicate the 95\% confidence intervals.

Compared to the reference scenario where all UX criteria are at their lowest level, the results show that increasing each of them has a statistically significant positive effect on the probability that the prototype screen is selected.\footnote{The ability to stop or pause an agent, however, barely fails to reach the 95\% significance level ($p = 0.0502$) and is only statistically significant at the 90\% confidence level.} We can also see that the highest level of agent transparency has the strongest impact. Changing agent transparency from low to high increases the probability that responses prefer this prototype screen by $37.79$ percentage points with a 95\% confidence interval within [$31.93$; $41.94$]. Increasing the human accountability UX criteria by adding a warning message to the low confidence recommendations changes the selection probability by $22.46$ percentage points [$15.92$; $28.17$]. In contrast, allowing users to stop or pause the agent workflow only increases the selection probability by $5.11$ percentage points [$0.00$; $10.12$].

We further assess the robustness of these findings in several ways. First, we address the statistical problem of multiple hypothesis tests by using the Bonferroni correction as well as adaptive shrinkage \citep{liu2023}. While Bonferroni correction renders the AMCE of the pause criteria insignificant, all other estimates remain unaffected. Second, we account for possible fatigue effects which could arise if respondents are getting tired after performing many choice tasks by restricting the sample to the first six choice tasks per respondent. We also estimate interactions between the task number and each individual AMCE as suggested by \citep{jenke2021}. Again, the results remain stable across these alternative specifications. Finally, we exclude responses based on response time. Specifically, we exclude the fastest ten percent and the slowest ten percent of responses and re-estimate the conditional logistic regression model. While this specification causes the effect of the stop/pause agents criterion to become insignificant, the other AMCEs remain statistically significant and substantively comparable in their effect sizes.

Due to the selection of the sample, inferences from this experiment are only valid for the specific population of users already working with AI tools in a business context. Although the sample size is sufficient to identify general effects, the identification of heterogeneous treatment effects across subgroups requires a larger sample and we leave this task to future research.\footnote{A commonly used heuristic to determine the minimally required sample size $n$ is given by the following inequality: $n \geq 500\frac{c}{a \times t}$, where $t$ is the number of choice tasks one respondent performs, $a$ is the number of alternatives per choice task, and $c$ is the largest number of levels for any of the UX criteria. In our case, the inequality is satisfied as we have $107 \geq 62.5$. Yet, it is important to note that this rule-of-thumb neither accounts for the desired precision nor power level. It also does not define the optimal sample size but rather an easy formula to determine the bare minimum during the design phase of a study \citep{assele2023,orme2019,johnson2003}.}

Taken together, the conjoint experiment provides robust empirical evidence for the importance of the identified UX criteria. While the ability to interrupt agent workflows only has a minor impact on the choices, highlighting human accountability and especially providing detailed information on the agent workflow, including reasoning, next actions, and details on data sources strongly increases the likelihood that respondents prefer the prototype screen. Hence, although there is a threat of information overload, this study provides empirical evidence that transparency in the agentic workflow has the strongest impact on user preferences.

\section{Discussion}
This study examined what constitutes a positive user experience with AI agents in enterprise workflows, identifying eight UX principles and underlying criteria validated through multiple methods (see Table~\ref{tab:overview}).

While HCAI and UX frameworks provide essential conceptual scaffolding for human-centered AI \citep{xu2026,shneiderman2020,xu2025,xu2024}, the present work extends this literature by deriving and prioritizing UX principles that move from general design philosophy toward measurable, product-relevant guidance for human-AI agent interaction in business contexts and therefore into a domain-specific UX framework for enterprise human-agent interaction.

{\bfseries Human Control.}
Consistent with Bradshaw et al.'s emphasis on interpredictability and directability \citep{bradshaw2011}, we identified in this study that human control emerges as the dominant principle. Participants understood control as meaningful oversight at points of business impact rather than micromanagement, aligning with governance-oriented views that emphasize human-in-command regimes and clear liability assignment \citep{xu2026,kolt2025}. Our findings are aligned with the current literature, in which agents are framed as powerful assistants that must remain embedded in existing socio-technical structures of responsibility rather than acting as autonomous decision-makers \citep{xu2025,xu2024}. The applied conjoint experiment yield significant results, extending HCAI work by showing how human control is concretely implemented in design via agent transparency, human accountability and pause/stop mechanisms.

{\bfseries Reliability, Safety, and Robustness.}
The high prioritization of this principle by industry professionals reflects broader trustworthy-AI literature treating safety as a prerequisite for adoption rather than a differentiator \citep{xu2026,shneiderman2020}. The characterization of unreliable agents as operationally disruptive due to inconsistent or misleading outputs reinforces findings from security-oriented surveys \citep{deng2025}. Our study adds interaction-level detail by highlighting the importance of visible data recency, confidence indicators, and risk cues in everyday enterprise use.

{\bfseries Data Privacy and Governance.}
Reflecting concerns raised in the NIST AI RMF \citep{NIST2023} and cross-industry trustworthiness reviews \citep{nastoska2025}, data privacy ranked highly in our study. Participants described data misuse as catastrophic and often irreversible, especially in regulated sectors, while noting that overly restrictive controls limit usefulness—echoing trade-off discussions balancing privacy, fairness, and utility \citep{nastoska2025}. Our findings demonstrate that least-privilege access based on role, legible denial behaviors, strict privacy protection, and sensitive data handling must be considered as interaction-level design concerns, not merely as elements of the underlying infrastructure.

{\bfseries Context Awareness.}
In line with HCAI’s emphasis on socio-technical context and Xu’s UX 3.0 focus on ecosystem-level experiences, our study pointed out context-awareness as a key differentiator between generic chatbots and genuinely useful enterprise agents \citep{xu2026,xu2025,xu2024}.

{\bfseries Transparency and Explainability.}
Consistent with explainability work at NIST and in applied HCI \citep{NIST2023,quinn2020}, our study showed that transparency was seen as essential for verification and control. Crucially, transparency needs proved dynamic: Detailed explanations are critical for high-stakes tasks and novice users, but experienced users increasingly prioritize speed once trust is established. This nuance extends existing frameworks by providing concrete evidence for graduated transparency patterns in enterprise agent use.

{\bfseries Ecosystem Integration and Collaborative Partnership.}
Echoing governance analyses \citep{kolt2025} and HCAI-MF’s multi-level design paradigms \citep{xu2026}, business users and industry professionals preferred agents are embedded into existing tools and data systems, orchestrate complex workflows, and behave like collaborative partners while respecting permissions and approval boundaries.

{\bfseries Responsiveness and Intuitive Interaction.}
In this study, this principle emerged as necessary for adoption but was rarely prioritized over risk, accountability, and trust—consistent with HCI findings that basic usability is assumed in professional systems while higher-order qualities drive differentiation in AI-intensive contexts \citep{xu2024,zhang2005}. Participants noted they would abandon agents that are slow or confusing regardless of underlying model capability.

These findings demonstrate that positive experiences with AI agents in enterprise settings require human control, safety, privacy, and contextual integration to be treated as first-class design requirements, with additional principles serving as essential complements in practice.

\section{Conclusion}
This study contributes significantly to the Human-Computer Interaction and Human-Centered AI discussion providing an empirically grounded enterprise perspective on human-AI agent interaction with implications for both UX design and the development of AI agents.

The employed methodology allows us to identify and validate UX principles and underlying criteria that lead to a positive experience with AI agents in organizational contexts. The conjoint experiment demonstrates that higher levels for the three human control UX criteria evaluated have a statistically significant effect on user preference. Agent transparency exhibits the strongest impact in the selected business scenario, indicating it is the most critical aspect for product development teams to address. This underscores the business users' need to follow agent reasoning, understand task status, next steps, and data sources used. This helps gaining sufficient detail to assess whether the agent is meaningfully supporting their work.

The conjoint experiment further shows that business users prefer a low-confidence indicator paired with a warning message and accountability signal for decision-making over a standard disclaimer lacking both warning and accountability cues. While the stop/pause mechanism reaches statistical significance only at the 90\% confidence level, it is possible that the function of this button in the hypothetical workflow was not clear enough for our respondents which would be a possible explanation for this finding. Compared to the other two UX criteria, the ability to pause an agent has the smallest impact on user preferences.

Collectively, these findings demonstrate what users in a business context require to retain meaningful control over AI agents and how this translates into UX design decisions.

Across all study phases, our methodological objective was not only to understand what users expect from agents, but also to derive validated, implementable guidance that helps UX and engineering teams design interactions that preserve human control, foster trust and safety, respect organizational constraints such as privacy and governance, and ultimately create positive and effective experiences with AI agents in business settings.

This study does not account for potential effects of participant’s location, industry, company size, role, or experience level with AI agents. Participants in both the participatory design workshop and paper-and-pencil survey were self-selected through attendance at an event, which focused on AI agents in the enterprise world. Throughout the study, participants exhibited varying levels of experience with AI agents. Additionally, the prototype screens used in the conjoint experiment adhered to company-specific design guidelines, which constrained the representational choices for the selected UX criteria. The external validity of the findings further depends on the assumption that respondents understand the hypothetical scenario and the differences in the prototype screens (e.g., the function of the `pause' button).

Future research is needed to explore the possibility of heterogeneous treatment effects of the UX criteria on specific subgroups. Furthermore, exploring alternative implementations of the UX criteria into specific design elements and evaluate their impact on user preferences is a promising direction for additional research. As a next step, we plan to conduct a subsequent conjoint experiment focusing on data privacy and data governance as one of the identified top principles.

%%%%%%%%%%%%%%%%% ACKNOWLEDGEMENT %%%%%%%%%%%%%%%%%%%%%

\section*{Acknowledgement}
We would like to thank Giannis Misiakos , Carolin Achtermann, Lara Valenti, Jennifer Shore, Ana Tarasova, Carsten Schmitt, Diego Ferrin, Martin Schrepp, Manvi Verma, Marina Kellermann, and Klaus H\"auptle for their contribution to this research project.



\bibliographystyle{mnras}
\bibliography{bibliography}

%%%%%%%%%%%%%%%%% APPENDICES %%%%%%%%%%%%%%%%%%%%%

%%%%%%%%%%%%%%%%%%%%%%%%%%%%%%%%%%%%%%%%%%%%%%%%%%

\label{lastpage}
\end{document}